\documentclass[aps,prl,twocolumn,superscriptaddress,floatfix,preprintnumbers]{revtex4}

\newcommand{\bea}{\begin{eqnarray}}
\newcommand{\eea}{\end{eqnarray}}
\newcommand{\be}{\begin{equation}}
\newcommand{\ee}{\end{equation}}
\newcommand{\mw}   {\mbox{$m_{W}$}}
\newcommand{\mwsq}   {\mbox{$m_{W}^2$}}

\newcommand{\mz}   {\mbox{$m_{Z}$}}
\newcommand{\mzsq}   {\mbox{$m_{Z}^2$}}

\newcommand{\muf}{\mu_F}

\usepackage{amsmath}
\usepackage{graphicx}
\usepackage{xcolor}
\usepackage[normalem]{ulem}
\usepackage[utf8]{inputenc}

\begin{document}

\preprint{CERN-TH-2019-154, OUTP-19-13P, TIF-UNIMI-2019-17}

\title{NNLO QCD$\times$EW corrections to Z production in the $q\bar{q}$ channel}



\author{Roberto Bonciani}
\email[]{roberto.bonciani@roma1.infn.it}
\affiliation{Universit\`a di Roma ``La Sapienza'' and INFN Sezione di Roma1}

\author{Federico Buccioni}
\email[]{federico.buccioni@physics.ox.ac.uk}
\affiliation{Rudolf Peierls Centre for Theoretical Physics, 
Clarendon Laboratory, Parks Road,
Oxford OX1 3PU,
UK}

\author{Narayan Rana}
\email[]{narayan.rana@mi.infn.it}
\affiliation{INFN Sezione di Milano, Via Celoria 16, 20133 Milano, Italy}

\author{Ilario Triscari}
\email[]{ilario.triscari@unimi.it}
\affiliation{Dipartimento di Fisica ``Aldo Pontremoli'', University of Milano, Via Celoria 16, 20133 Milano, Italy}

\author{Alessandro Vicini}
\email[]{alessandro.vicini@mi.infn.it}
\affiliation{TH Department, CERN 1 Esplanade des Particules, Geneva 23, CH-1211, Switzerland}
\affiliation{Dipartimento di Fisica ``Aldo Pontremoli'', University of Milano, Via Celoria 16, 20133 Milano, Italy}
\affiliation{INFN Sezione di Milano, Via Celoria 16, 20133 Milano, Italy}


\date{\today}

\begin{abstract}
  We present the first results for the ${\cal O}(\alpha\alpha_s)$ corrections
  to the total partonic cross section of the process $q\bar q\to Z+X$,
  with the complete set of contributions, that include photonic and massive weak gauge boson effects.
  The results are relevant for the precise determination of the hadronic
  $Z$ boson production cross section.
  Virtual and real corrections are calculated analytically using the reduction
  to the master integrals
  and their evaluation through differential equations.
  Real corrections are dealt with using the reverse-unitarity method.
  They require the evaluation of a new set of two-loop master integrals,
  with up to three internal massive lines.
  In particular, three of them are expressed in terms of elliptic integrals.
  We verify the absence, at this perturbative order, of initial state mass singularities
  proportional to a weak massive virtual correction to the quark-gluon splitting.

\end{abstract}


\maketitle


The production of an electrically neutral gauge boson at hadron colliders
is one of the historical processes for our understanding of Quantum Chromodynamics (QCD).
The case of the decay of the $Z$ boson into a pair of high transverse momentum leptons
is known as Drell-Yan (DY) process and it is particularly important
for the setting of several high-precision tests of the electroweak (EW) sector
of the Standard Model (SM).
It allows for instance a precise measurement of the weak mixing angle and
of the properties of the $Z$ boson.
The $Z$ boson DY production  is one of the processes known with high perturbative accuracy. 
The pioneering calculations of the next-to-leading order (NLO) \cite{Altarelli:1979ub}
and next-to-next-to-leading order (NNLO) \cite{Hamberg:1990np} QCD corrections
to the total inclusive cross section
have been extended later to the fully differential description of the leptonic final state
\cite{Anastasiou:2003ds,Melnikov:2006kv,Catani:2009sm,Gavin:2010az}.
Finally, the evaluation of the next-to-next-to-next-to-leading order (N$^3$LO)
QCD corrections at the production threshold has been presented in
Refs.~\cite{Ahmed:2014cla,Ahmed:2014uya,Li:2014bfa,Catani:2014uta}
for the gauge boson total cross section and rapidity distribution.
The impact of the NLO EW corrections,
studied in Refs.~\cite{Baur:2001ze,CarloniCalame:2007cd,Arbuzov:2007db,Dittmaier:2009cr},
is at the ${\cal O}(1\%)$ level as far as the total cross section is concerned
and it is comparable to that of the NNLO QCD contributions.
Kinematic distributions may receive additional enhancements in specific phase-space regions,
yielding corrections at the ${\cal O}(10\%)$ level or more.
Since the high-precision determination of EW parameters requires control
over the kinematic distributions in some cases at the per mille level
(cf. Refs.~\cite{Alioli:2016fum,CarloniCalame:2016ouw,Bozzi:2015hha}
for a discussion on specific examples),
the evaluation of the mixed QCD-EW corrections has emerged as necessary
for both the study of the gauge boson resonances and of the high mass/momentum tails
of the kinematic distributions \cite{Balossini:2008cs,Balossini:2009sa}.
First analytic results have been presented in
Refs.~\cite{Kilgore:2011pa,Dittmaier:2014qza,Dittmaier:2014koa,Dittmaier:2015rxo,Dittmaier:2016egk},
and compared with the approximations available via Monte Carlo simulation
tools \cite{Barze:2012tt,Barze:2013fru}:
while the bulk of the leading effects, separately due to QCD and QED corrections,
can be correctly evaluated for several observables,
the remaining sub-leading QED effects and the genuine QCD-weak corrections are still missing in these tools.
Furthermore, a realistic estimate of the theoretical uncertainties must account for several sources
of ambiguity related to the recipes used in the matching of separate results
for the QCD and EW contributions to the scattering amplitude.
For these reasons an exact calculation of the full set of ${\cal O}(\alpha\alpha_s)$ corrections
to the DY processes is desirable.
In Refs.~\cite{deFlorian:2018wcj,Cieri:2018sfk,Delto:2019ewv} the mixed QCD-QED corrections
to the total cross section and transverse momentum spectrum of an on-shell $Z$ boson
have been discussed.
The evaluation of all the Master Integrals (MIs) relevant to compute the full set
of QCD-EW mixed corrections to DY process (including off-resonance terms)
has been documented in Refs.~\cite{Bonciani:2016ypc,Heller:2019gkq}.

In this letter, we present the first results for the total inclusive cross section
of production of an on-shell $Z$ boson in the quark-antiquark partonic channel,
including the complete set of QCD-EW corrections of ${\cal O}(\alpha\alpha_s)$.
We retain the dependence on the massive states exchanged in the loops.
As a consequence of that, the calculation involves a set of two-loop phase-space integrals,
previously not available in the literature.
Their analytic expression will be presented in a forthcoming paper. 
We also have the occasion to check the infrared structure of the corrections up to NNLO level,
including the cases where a massive EW boson is exchanged.
We verify the absence of initial state mass singularities proportional to a weak massive virtual
correction to the quark-gluon splitting.

The calculation we are presenting in this letter is an important step towards
the evaluation of the full set of QCD-EW corrections to the hadronic cross section. 

%

\section{Theoretical framework}

The inclusive production cross section $\sigma_{tot}$ of a $Z$ boson at hadron colliders $(pp\to Z+X)$ can be written, using the factorization theorem, as 
\begin{align}
&\sigma_{tot} (\tau) = \sum_{i,j\in q,\bar q, g, \gamma}
\int {\rm d}x_1 {\rm d}x_2 \hat{f}_i (x_1) \hat{f}_j (x_2) \hat\sigma_{ij} (z) \, .
\label{eq:sigmatot-bare}
\end{align}
In Eq.~(\ref{eq:sigmatot-bare}), $\tau=\frac{m_Z^2}{S}$ and $z=\frac{m_Z^2}{\hat{s}}$ are the ratio of the squared $Z$ boson mass, $\mz$, with $S$ and $\hat{s}$, the hadronic and partonic center of mass energy squared, respectively. $S$ and $\hat{s}$ are related by $\hat{s}=x_1 x_2 S$ through the Bjorken momentum fractions $x_1, x_2$.
The bare cross section $\hat\sigma_{ij}$ of the partonic process  $ij\to Z+X$ is convoluted with the bare parton densities $\hat{f}_i (x)$. The sum over $i,j$ includes quarks ($q$), antiquarks ($\bar{q}$), gluons ($g$) and photons ($\gamma$). In the SM, we have a double expansion of the partonic cross sections in the electromagnetic and strong coupling constants, $\alpha$ and $\alpha_s$, respectively:
\begin{equation} 
\hat\sigma_{ij} (z) = \sum_{m,n=0}^{\infty}\alpha_s^m \alpha^n ~ \hat\sigma_{ij}^{(m,n)} (z) \, ,
\label{eq:sigmaexp}
\end{equation}
where $\hat\sigma_{ij}^{(m,n)}$ is the correction of ${\cal O}(\alpha_s^m\alpha^n)$ to the lowest-order inclusive total cross section $\hat\sigma_{ij}^{(0,0)}$ of the partonic scattering $ij \to Z$. For a given initial state, the inclusive total cross section receives contributions from processes with different final state multiplicities, due to real parton emissions. In this letter we focus on the $q \bar{q}$ initiated scattering, and, for definiteness, we treat the case of an up-type quark: $q \bar{q} = u \bar{u}$.
The full set of ${\cal O}(\alpha\alpha_s)$ corrections to $\hat\sigma_{u\bar{u}}$ stems from the evaluation of the following scattering processes:
\bea
&&u\bar u \to Z \, , \label{eq:proc1}\\
&&u\bar u \to Zg \, , \label{eq:proc2}\\
&&u\bar u \to Z\gamma \, , \label{eq:proc3}\\
&&u\bar u \to Zg\gamma \, , \label{eq:proc4}\\
&&u\bar u \to Z u\bar u \, , \label{eq:proc5}\\
&&u\bar u \to Z d\bar d \, , \label{eq:proc6}
\eea
where $d$ represents a down-type massless quark. 
Explicit expressions for the process (\ref{eq:proc4}) and QCD-QED contributions to process (\ref{eq:proc1})
have been presented in Ref.~\cite{Bonciani:2016wya} and Ref.~\cite{Kilgore:2011pa,H:2019nsw}, respectively.
The corresponding results for $d\bar d$ initiated subprocesses can be derived from our results with the replacements $Q_u\leftrightarrow Q_d$, $I_u^{(3)}\leftrightarrow I_d^{(3)}$, where $Q_f,\,I_f^{(3)}$ are the electric charge and the third component of the weak isospin, for a fermion $f$, respectively.

The process (\ref{eq:proc1}) receives contributions from two-loop $2 \to 1$ Feynman diagrams,
that have to be interfered with the tree-level $u \bar{u} \to Z$ and constitute
the virtual corrections.
The processes (\ref{eq:proc2})--(\ref{eq:proc3}) receive contributions
from one-loop $2 \to 2$ Feynman diagrams that have to be interfered
with the corresponding tree-level.
We refer to them as real-virtual corrections.
The last three processes, (\ref{eq:proc4})--(\ref{eq:proc6}),
receive contributions from tree-level $2 \to 3$ Feynman diagrams interfered
with themselves and we refer to them as double-real corrections.

The full set of ${\cal O}(\alpha\alpha_s)$ corrections can be organized
in two gauge invariant subsets: QCD-QED and QCD-weak contributions.
Processes (\ref{eq:proc1})--(\ref{eq:proc5}) contribute to the former,
and processes (\ref{eq:proc1}), (\ref{eq:proc2}), (\ref{eq:proc5}) and (\ref{eq:proc6}) to the latter.
While one gluon exchange, real or virtual, is always present,
we identify three groups of contributions to the amplitudes depending
on the presence of one real or virtual photon, of one virtual $Z$ boson,
or of one/two virtual $W$ bosons.
We further observe that the last two groups are separately gauge invariant.
In our definition of total cross section we do not include the processes
with the emission of one extra massive on-shell gauge boson, as their measurement
depends on the details of the experimental event selection.
Furthermore, these corrections do not contribute to the infrared structure of the process.

The amplitude of the two tree-level processes (\ref{eq:proc5}) and (\ref{eq:proc6})
has two components of ${\cal O}(\sqrt{\alpha} \alpha_s)$ (an internal gluon exchange)
and ${\cal O}( \sqrt{\alpha} \alpha )$ (an internal weak boson exchange),
respectively and their interference is, therefore, of ${\cal O}(\alpha^2 \alpha_s)$.

\section{Computational details}

We follow a diagrammatic approach to obtain all the relevant contributions
to the inclusive production cross section $u\bar{u} \to Z+X$.
A detailed description of the computation will be presented in a dedicated publication.
In this letter we sketch an outline of the procedure.
We need to include contributions with two-loop virtual corrections,
with one real emission and one loop (real-virtual), with two real emissions (double-real),
and factorisable contributions stemming e.g. from the interference of two one-loop diagrams.
We treat all the processes with the same algorithmic approach.
Firstly, we compute all the Feynman diagrams contributing to a given amplitude with {\texttt{FeynArts}}
\cite{Hahn:2000kx} and {\texttt{QGRAF}} \cite{Nogueira:1991ex},
we perform algebraic simplifications with {\texttt{FORM}} \cite{Vermaseren:2000nd}
and {\texttt Mathematica};
we use integration-by-parts (IBP) \cite{Tkachov:1981wb,Chetyrkin:1981qh,Laporta:2001dd} and Lorentz-invariance (LI) identities \cite{Gehrmann:1999as} to reduce the Feynman integrals to MIs.
The reduction to the MIs is carried out using the computer programs
\texttt{Kira} \cite{Maierhoefer:2017hyi},
\texttt{LiteRed} \cite{Lee:2012cn,Lee:2013mka}, and
\texttt{Reduze 2} \cite{Studerus:2009ye,vonManteuffel:2012np}
The entire procedure is performed within dimensional regularization in $D=4-2 \varepsilon$ space-time dimensions.
%
%
Then, we employ the method of differential equations
\cite{Kotikov:1990kg,Remiddi:1997ny,Gehrmann:1999as,Argeri:2007up,Henn:2014qga}
to compute the MIs, for both the pure virtual and real emission corrections.
In the latter case, the phase-space delta functions are dealt with via the reverse unitarity technique
\cite{Anastasiou:2002yz,Anastasiou:2012kq},
which is based on the observation that the replacement known as Cutkowsky rule
holds in terms of distributions: 
\begin{equation}
\delta(p^2-m^2) \rightarrow 
\frac{1}{2\pi i} \left( \frac{1}{p^2-m^2+i\eta} - \frac{1}{p^2-m^2-i\eta} \right). 
\end{equation}
It is thus possible to rewrite the phase-space measure of each final-state particle
as the difference of two  propagators with opposite prescriptions for their imaginary part
(where $\eta$ stands for an infinitesimal positive real number).
We transform the integration over the full phase space of the additional parton/s for processes
(\ref{eq:proc2})--(\ref{eq:proc6}), into the evaluation of the cut two-loop integrals with an on-shell condition on the lines that correspond to the final-state particles.

The pure virtual MIs are already available in the literature
\cite{Aglietti:2003yc,Aglietti:2004tq,Aglietti:2004ki,Aglietti:2007as,Bonciani:2010ms,Kotikov:2007vr},
in the case of off-shell $Z$ boson.
Since in our case the $Z$ boson is on-shell,
we have computed these integrals taking the appropriate on-shell limit.
The two- and three-body phase-space MIs with only gluon or photon lines are already
available in the literature \cite{Anastasiou:2012kq}.
To validate our routines developed for the present calculation, however,
we have recomputed them and found complete agreement with the known expressions.
We have computed all the new MIs, with one or two internal massive lines,
with the differential equations method. 
We have fixed the boundary conditions calculating the soft limit ($z\to 1$) of the MIs.

After integration over the phase-space of the emitted real partons,
the partonic total cross section depends solely on the variable $z$.
The virtual contributions are proportional to $\delta(1-z)$
and are therefore constants,
  which are found from the on-shell limit of the virtual MIs,
  i.e. evaluating the corresponding generalised harmonic polylogarithms (GPLs)
\cite{Goncharov:polylog,Goncharov2007,Remiddi:1999ew,Vollinga:2004sn}
  at $z=1$.
  All the constants arising from this limit can be reduced to the basis
  introduced in Ref.~\cite{Henn:2015sem}.
The part that corresponds to processes (\ref{eq:proc2})--(\ref{eq:proc6})
is expressed almost entirely in terms of $\delta(1-z)$ and of GPLs,
or cyclotomic Harmonic Polylogarithms \cite{Ablinger:2011te}, functions of $z$.
In some parts of the calculation
the package {\texttt{HarmonicSums}} \cite{Ablinger:2013cf} has been used.
Three MIs appearing in processes (\ref{eq:proc5}) and (\ref{eq:proc6})
satisfy elliptic differential equations,
whose homogeneous behaviour has already been studied in Ref.~\cite{Aglietti:2007as}.
We have obtained their complete solution with a series expansion around $z=1$
(see for instance \cite{Pozzorini:2005ff,Aglietti:2007as,Lee:2017qql,Lee:2018ojn,Bonciani:2018uvv}).

In the calculation of the MIs, the masses of the $W$ and $Z$ bosons are set equal to $\mz$,
to avoid the presence of an additional energy scale in the problem,
which would make the analytical solution of the differential equations in terms of known functions
more complicated.
While for the virtual corrections this choice is not strictly necessary,
since the knowledge of the MIs for off-shell $Z$ would allow
for a complete and exact calculation in the case of $m_W \not = m_Z$,
the reduction of one mass scale in the computation of the real emission processes is in fact very effective
and reduces the complication of the calculation.
Moreover, the equal-mass choice does not prevent us from obtaining an analytical solution
with arbitrary precision for each of the affected MIs.
In fact, we can perform an expansion of the integrand in powers of
the ratio  $\delta_m^2=(\mzsq-\mwsq)/\mzsq$,
and reduce all the terms of the series to a combination of the same basic equal-mass MIs.
We stress that the couplings of the $Z$ boson to fermions are expressed in terms of the physical
value of the weak mixing angle $\sin^2\theta_W=1-\mwsq/\mzsq$.

\subsection{Ultraviolet renormalization}

The calculation is performed in the EW background field gauge (BFG) \cite{Denner:1994xt}, which allows the identification of two sets of ultraviolet (UV) finite amplitudes. On the one hand, the combination of 1PI vertex and external quark wave function corrections, which satisfies, also in the EW SM, a QED-like Ward identity, with the consequent cancellation of the UV poles. On the other, the external $Z$ boson wave function and the lowest-order coupling renormalization corrections, whose combination, order-by-order in perturbation theory, is also UV finite.

We need to perform the renormalization of the couplings and the fields up to ${\cal O}(\alpha\alpha_s)$ for process (\ref{eq:proc1}), while we need only the ${\cal O}(\alpha)$ renormalization of process (\ref{eq:proc2}).
One-loop QCD corrections to processes (\ref{eq:proc1}) and (\ref{eq:proc3})  are UV finite, after field renormalization, again because of a QED-like Ward identity. We remark that the $Z$ boson field and the EW couplings do not receive ${\cal O}(\alpha_s)$ renormalization corrections.
The renormalization of the quark field receives EW corrections and we consider this in the on-shell scheme.
The EW gauge sector of the SM Lagrangian depends on three parameters $(g,g',v)$, the two gauge couplings and the Higgs-doublet vacuum expectation value. After the introduction of counterterms and renormalized parameters,
we express the latter as a combination of $(G_\mu,\mw, \mz)$
\footnote{
  The counterterm necessary in the case we replace the Fermi constant with the fine structure
  constant is described in Ref.~\cite{Degrassi:2003rw}.
  An alternative scheme where the effective leptonic weak mixing angle appears as input parameter
has been discussed in Ref.\cite{Chiesa:2019nqb}.
},
respectively the Fermi constant, the $W$ and $Z$ boson masses.

A subset of the EW corrections can be reabsorbed in a redefinition of the weak mixing angle that appears in the vector coupling of the $Z$ boson to fermions.
These corrections are split, in the EW BFG, in two UV-finite groups, one due to vertex corrections, the other due to external $\gamma-Z$ corrections and to the weak mixing angle counterterm (a shortcut for a combination of $W$ and $Z$ mass counterterms). In BFG the second group vanishes, because of a Ward identity \cite{Denner:1994xt}
satisfied by the $\gamma-Z$ wave-function correction.

\subsection{Infrared singularities and mass factorization}

The ${\cal O}(\alpha\alpha_s)$ corrections are organized in two gauge invariant subsets: QCD-QED and QCD-weak contributions. The former involve the exchange of two massless bosons, yielding the maximal degree of infrared singularity at the second perturbative order, i.e. $\varepsilon^{-4}$. The latter have only the poles due to a soft and/or collinear gluon.
The cancellation of the soft singularities takes place separately in the two subsets, once the contribution of virtual corrections and of the corresponding soft real emissions are combined. To be more precise, for the QCD-QED subset, the process (\ref{eq:proc5}) does not yield soft singularities, so that the cancellation takes place when the processes (\ref{eq:proc1})--(\ref{eq:proc4}) are combined. In the case of the QCD-weak subset, soft singularities appear only in processes (\ref{eq:proc1}) and (\ref{eq:proc2}) and cancel when the two are summed. When we consider the combination of the cross sections of the processes (\ref{eq:proc1})--(\ref{eq:proc6}) we are thus left with initial state collinear singularities only. The processes (\ref{eq:proc1})--(\ref{eq:proc5}) contribute to initial-state collinear singularities within the QCD-QED subset, while in the QCD-weak case only processes (\ref{eq:proc1})--(\ref{eq:proc2}) have initial state collinear singularities of QCD origin. These singularities can be removed by mass factorization. The physical parton densities $f_i(x,\muf)$ are defined, at the factorization scale $\muf$, by introducing the mass factorization kernel $\Gamma_{ij}$, which subtracts
the initial state collinear singularities
\begin{equation}
\hat{f}_i = f_j \otimes \Gamma_{ij} \,.
\label{eq:baretophysicalPDF}
\end{equation}
The kernel can be expanded as a series in $\alpha$ and $\alpha_s$
\begin{equation}
\Gamma_{ij}
=
\sum_{m,n=0}^{\infty}\alpha_s^m \alpha^n \Gamma_{ij}^{(m,n)}\,,
\label{eq:gammaexp}
\end{equation}
where $\Gamma_{ij}^{(1,0)}$ is the QCD leading order (LO) splitting kernel, $\Gamma_{ij}^{(0,1)}$ is its QED analogue and $\Gamma_{ij}^{(1,1)}$ is the mixed QCD-QED contribution to the splitting kernels, recently presented in Ref.~\cite{deFlorian:2015ujt}. 
After the replacement of Eq.~(\ref{eq:baretophysicalPDF}) in Eq.~(\ref{eq:sigmatot-bare}), 
we obtain the total cross section expressed in terms of subtracted, finite, partonic cross sections $\sigma_{ij}(z, \muf)$:
\be
\sigma_{tot}(z) =\!\!\!\!\!\!
\sum_{i,j\in q,\bar q, g, \gamma}
\!\!\!\!\!\!
\int {\rm d}x_1 {\rm d}x_2  
f_i(x_1,\muf) f_j(x_2,\muf) \sigma_{ij} (z, \muf)\, .
\ee
The $\sigma_{ij}$ admit a perturbative expansion in powers of $\alpha$ and $\alpha_s$, in analogy to Eq.~(\ref{eq:sigmaexp}). 
In this letter, we present the results for $\sigma_{u\bar{u}}^{(1,1)}$.

In processes (\ref{eq:proc1}) and (\ref{eq:proc2}) the weak virtual correction to the splitting vertex $q\to qg$
might induce an additional contribution to the subtraction kernel $\Gamma_{ij}^{(1,1)}$. 
However, we have checked that such a term vanishes, in the massless quark case, as a consequence of the conservation of the vector and axial-vector currents.

\section{Results}

In order to discuss the size of the different sets of radiative corrections,
we define:
\bea
\alpha_s\alpha \sigma_{u\bar{u}}^{(1,1)}
\!\!\!\!&=&\!\!
\sigma_{u\bar{u}}^{(0)}
\left(
\Delta_{u\bar{u},\gamma}^{(1,1)}
+
\Delta_{u\bar{u},Z}^{(1,1)}
+
\Delta_{u\bar{u},W}^{(1,1)}
\right)\,\,\,\,\,\,\,
\eea
where $\sigma_{u\bar{u}}^{(0,0)} \equiv \sigma_{u\bar{u}}^{(0)} \delta(1-z) =
4\sqrt{2}  G_\mu (\pi/N_c) (C_{v,u}^2+C_{a,u}^2) \delta(1-z)$
is the Born cross section of the process $u\bar u\to Z$, with $N_c$ the number of colours
and $C_{v/a,u}$ the vector/axial-vector couplings of the $Z$ boson to the up quark.
$\Delta_{u\bar{u},K}^{(1,1)}$ with $K=\gamma,Z,W$ are the corrections due to the exchange
of a photon, a $Z$ boson, and of one or two $W$ boson/s including the lowest order charge renormalization counterterms,
respectively.
For the sake of comparison, we introduce the NLO-QCD correction to the same partonic process,
defined as $\alpha_s \sigma_{u\bar{u}}^{(1,0)} = \sigma_{u\bar{u}}^{(0)} \,\, \Delta_{u\bar{u}}^{(1,0)}$.
\begin{figure}[ht]
  \includegraphics[width=0.47\textwidth]{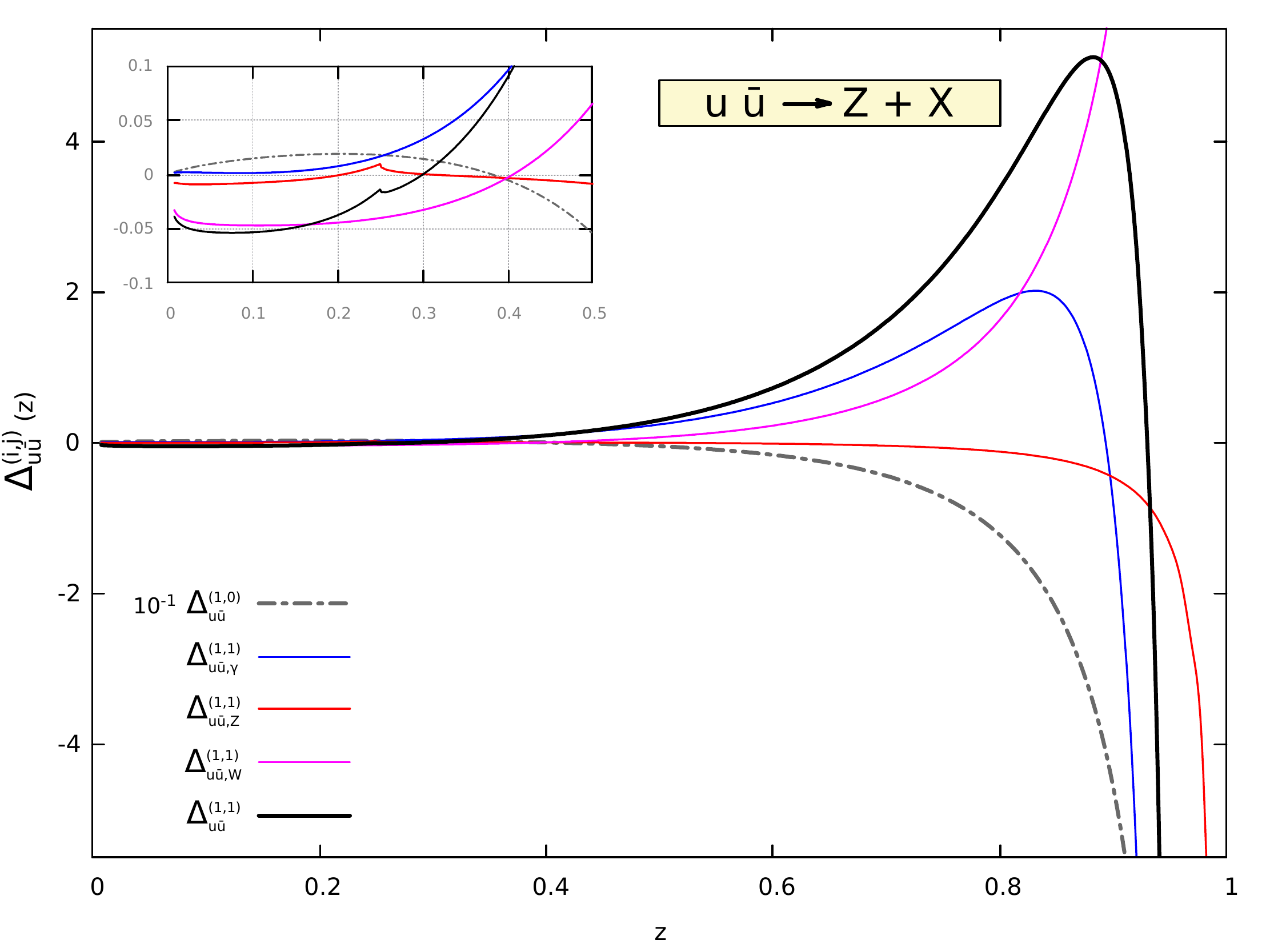}
  \caption{Corrections factors $\Delta_{u\bar{u}}^{(1,0)}$
    and $\Delta_{u\bar{u},K}^{(1,1)}$ with $K=\gamma,Z,W,ct$,
    as a function of the partonic variable $z$.
    The NLO-QCD correction $\Delta_{u\bar{u}}^{(1,0)}$ (grey dashed)
    is divided by a factor 10.
    We show the ${\cal O}(\alpha \alpha_s)$ total correction (black solid),
    and the contribution of the different subsets $K=\gamma,Z,W$
    in blue, red, and magenta, respectively.
    \label{fig:deltas}}
\end{figure}
In Figure \ref{fig:deltas} we present, as a function of the partonic variable $z$,
the contribution of the different subsets of diagrams, $\Delta_{u\bar{u},K}^{(1,1)}$ with $K=\gamma,Z,W$, and their sum.
We also plot $\Delta_{u\bar{u}}^{(1,0)}$, divided by a factor 10.
We exclude from the plot all the
contributions proportional to $\delta(1-z)$,
while we keep all the plus-distribution terms, limiting the plot at $z=0.99$.
For the numerical evaluation, we use the following input parameters:
$m_W=80.385$ GeV, $m_Z=91.1876$ GeV, $G_\mu=1.1663781\times 10^{-5}$ GeV$^{-2}$,
$m_t=173.5$ GeV, $m_H=125$ GeV, $\alpha_s(\mz)=0.118$.
$m_t$ and $m_H$ are the top quark and Higgs boson mass, respectively.
We set the factorisation scale $\mu_F=\mz$.

We observe that, in the high-energy limit ($z\to 0$), the cross sections
are damped by the incoming flux factor, proportional to $z$.
The divergent behaviour for $z\to 1$, due to the exchange of at least one massless boson,
is also evident for all the contributions.
The values of the EW charges, in the two subsets with one $Z$ (red) or with one/two $W$s exchange (magenta),
are responsible for the different size and for the opposite sign of the two contributions,
visible in the $z\to 1$ limit.
We observe that in the case of the $d\bar d\to Z+X$ process,
the contributions with one/two $W$s exchange have
similar size but opposite sign.
The total contribution to the hadron-level cross section
from this subset of diagrams of the two partonic processes
is expected to undergo an important cancellation,
modulated by the convolution with the proton PDFs.
The QCD-QED corrections, shown in blue in Figure \ref{fig:deltas}, are not monotonic,
contrary to the NLO-QCD ones and have a maximum for $z\sim 0.85$.
They are smaller than the QCD-weak contribution for $z\in [0.8,0.9]$,
but become larger in absolute size when $z\to 1$,
because of the higher power of the threshold logarithms.
The possibility of having a second $Z$ boson in a resonant configuration
yields the kink of the $\Delta_{u\bar{u},Z}^{(1,1)}$ curve (red) at $z=1/4$,
as it can be observed in the inset of Figure \ref{fig:deltas}.

In conclusion,
we have presented the first results for the total inclusive partonic cross section
for the process $q\bar q\to Z+X$, including the exact ${\cal O}(\alpha \alpha_s)$ corrections,
with both photon and $W/Z$ boson exchanges.
The results are analytic and are expressed in terms of GPLs,
but also contain three elliptic MIs, which have been computed with a series expansion around $z=1$.
The complete solution of the infrared structure of the process
and the exact evaluation of all the relevant virtual corrections
represent an important step towards the evaluation of the hadron-level cross section
for $Z$ production at this perturbative order.

\begin{acknowledgments}
{\bf Acknowledgments.}
  N.R. thanks the CERN Theory Department for hospitality and support during the completion of this work.
  A.V. is supported by the European Research Council
  under the European Unions Horizon 2020 research and innovation Programme (grant agreement number 740006).
F.B. warmly thanks the Physics Institute of the University
of Zurich where large part of this work was carried out and
acknowledges support from the 
Swiss National Science Foundation (SNF) under contract BSCGI0-157722. 
The research of F.B. was partially supported by the
ERC Starting Grant 804394 {\sc{hip}QCD}. 
R.B. and N.R. acknowledge the COST (European Cooperation in Science and
Technology) Action CA16201 PARTICLEFACE for partial support.
\end{acknowledgments}

\bibliography{BBRTV}

\end{document}